\documentclass{PoS}
\newcommand{\nn}{\nonumber}

\title{Magnetic Monopoles in (Noncommutative) Quantum Mechanics}

\ShortTitle{Magnetic Monopoles in (NC) QM}

\author{\speaker{Samuel Kov\'a\v{c}ik}\\
        Dublin Institute for Advanced Studies, 10 Burlington Road, Dublin 4, Ireland\\
        E-mail: \email{skovacik@stp.dias.ie}}
        
        \author{Peter Pre\v{s}najder \\
	    Faculty of Mathematics, Physics and Informatics, Comenius University Bratislava, Mlynsk\'a dolina, Bratislava, 842 48, Slovakia\\
        E-mail: \email{presnajder@fmph.uniba.sk}}

\abstract{We utilize the close relation between the complex space $\textbf{C}^2$ and the real space $\textbf{R}^3$ to reformulate quantum mechanics in a manner which allows to, either or both, describe magnetic monopoles and quantize the underlying space, obtaining (noncommutative) quantum mechanics (with magnetic monopoles).\\

Ref. number: DIAS-STP-18-08}

\FullConference{Corfu Summer Institute 2017 'School and Workshops on Elementary Particle Physics and Gravity'\\
		2-28 September 2017\\
		Corfu, Greece}

\begin{document}

\section{Introductions}
We present an alternative description of magnetic monopoles in quantum mechanics (QM) defined on either commutative or noncommutative (NC) space, in both cases by generalizing the considered class of physical states. As the reader might be unfamiliar with at least one of the topics, we will introduce them both.
\subsection{Magnetic monopoles}
Magnetic monopoles have never been observed directly, yet they keep appearing in physical theories for nearly a century. The only evidence of their existence, even though just indirect, is the discreteness of electric charge, which is required by the consistency of the theory and is indeed observed in nature. The general belief is that magnetic monopoles do exist but are too heavy (on the grand unification scale) to be produced in accelerators and those formed shortly after the Big bang were diluted by the process of inflation. 

Formalism of the classical electromagnetism is largely build on the premise that magnetic monopoles do not exist, instead of working with electromagnetic fields we work with electromagnetic potentials, the non-existence of magnetic monopoles follows directly from $\mbox{div }\textbf{B} = \mbox{div rot }\textbf{A} = 4\pi \rho_M = 0$ valid for regular $\textbf{A}$. Obviously, to describe monopoles we could just return to the notion of fields, yet from the point of view of theories such as special relativity or quantum mechanics, the notion of potentials is largely preferred. 

Fortunately, there is a workaround. Dirac showed that sources of magnetic fields can be described by potentials singular on a (half-)line, so-called Dirac string, see \cite{dirac}. Later, Yang found that the singular behavior is not necessary. Instead, one can use two regular potentials defined in different but overlapping regions, each of them avoiding the singularity and together covering entire space outside the monopole, see \cite{Yang}. 

In modern theories, monopoles often appear as topological solutions to field theories (traces of topology can be found already in the approach of Yang). The notorious examples are the grand unification monopoles of 't Hooft and Polyakov \cite{hooft,pol} and the Kaluza-Klein monopoles of Gross, Perry and Sorkin \cite{Sorkin:1983ns, Gross:1983hb}.

We present an alternative description of magnetic monopoles by lifting the theory one dimension up, a method explored in the context of electromagnetism in \cite{Ryder, Minami}. However, our focus is on quantum mechanics, therefore our results are to be compared to those of Zwanziger, who studied such theory in some detail in \cite{zwanziger}.

\subsection{Noncommutative theories}
Noncommutative space is a space whose very close points cannot be distinguished. Its coordinates do not commute in a similar fashion as the momentum and the coordinate operators do not commute in ordinary QM. Such spaces sometimes serve as an effective description for physical theories on scales reachable in laboratory experiments. However, they are also a rather general consequence of theories of quantum gravity, where they become relevant only on the Planck scale. 

That the space cannot be probed below this scale follows from a simple thought-experiment: If one tries to distinguish two points separated by a Planck length, the used photon (or any other particle for that matter) would be energetic enough to be hidden under its own event horizon, a black hole would be formed and no information could be obtained. 

Consider the difference in the formalism of classical and quantum mechanics caused by the phase space noncommutativity of the latter. A similar leap is to be expected in the case of noncommutativity of the underlying space. The ultimate goal is to obtain a proper formulation of NC quantum field theory, as dual to UV distances, UV energies are removed as well. However, we settle for NC QM to be our starting point to understand the consequences of the space noncommutativity, to learn how to formulate the theory in a consistent way and to study new prospects (and problems) it brings.

There are many examples of NC spaces, see \cite{groenewold, snyder,fuzzy, Con}. We will consider the three-dimensional, rotationally invariant space $\textbf{R}^3_\lambda$, which was constructed in \cite{Jabbari} and studied in greater detail in \cite{GP1,GP2,vel, LRL, plane, mbh}. We have shown in  \cite{mm,mm2} that monopoles can be described in $\textbf{R}^3_\lambda$ QM by considering a generalized class of physical states. Even though the space $\textbf{R}^3_\lambda$ is usually understood to be a sequence of concentric fuzzy spheres, it can be also obtained by the means of Kontsevich quantization. Applying the same approach in the pre-quantized theory, one can describe magnetic monopoles in ordinary $\textbf{R}^3$ QM as well (as was studied in \cite{mmc} and will be briefly sketched in the next section).

The paper is organized as follows: in Section 2 we present an alternative description of magnetic monopoles in ordinary QM. In Section 3 we show a brief construction of QM in NC space $\textbf{R}^3_\lambda$. In Section 4 we discuss magnetic monopoles in NC QM introduced in a similar fashion to Section 2. The last Section is devoted to conclusions.
\section{Magnetic monopoles in quantum mechanics}
A quantum-mechanical theory usually requires two ingredients: a Hilbert space of states and physical observables realized as operators on it. The notorious example is the Hilbert space of square-integrable functions equipped with the norm
\begin{equation} \label{norm}
||\Psi || ^2 = \int \Psi^* (x) \Psi(x) d^3x ,
\end{equation}
with the position or the momentum operators defined as
\begin{eqnarray}
\hat{x}_i \Psi(x) &=& x_i \Psi(x), \\ \nn
\hat{p}_i \Psi(x) &=& - i \partial_i \Psi(x).
\end{eqnarray}
This is built on the underlying space $\textbf{R}^3$, however, it is not a necessity, the Hilbert space with the observables of the same structure can be built on $\textbf{C}^2$ as well. This space has two complex coordinates $z_\alpha, \alpha =1,2$ and is equipped with the Poisson structure
\begin{equation} \label{Pois}
\{ z_\alpha ,z^*_\beta \}\ =\ - i\,\delta_{\alpha\beta}\,,\ \  \{ z_\alpha ,z_\beta \}\ =\ \{ z^*_\alpha ,z^*_\beta \}\ =\ 0.
\end{equation}
We can define a Hilbert space of square-integrable functions $\Phi(z,z^*)$ equipped with the norm
\begin{equation} \label{norm1}
||\Phi^2 || = \int \frac{2 r}{\pi} \Phi^* (z,z^*) \Phi(z,z^*) dz d\bar{z},
\end{equation}
where $r = \bar{z} z$. The Poisson structure \ref{Pois} allows us to naturally construct many important operators, for example the Laplacian
\begin{equation} \label{lapl}
\Delta \Phi(z,z^*) =\frac{1}{r}\,\{z^*_\alpha, \{z_\alpha,\Phi(z,z^*)\}\}.
\end{equation}

So far this seems nothing like the ordinary QM. However, recall that the spaces in question, $\textbf{R}^3$ and $\textbf{C}^2$, are closely related as their rotational groups are locally isomorphic. In fact, one can parametrise $\textbf{C}^2$ as 
\begin{eqnarray}\label{euler}
z_1 =& \sqrt{r} \cos \left(\theta /2 \right) e^{\frac{i}{2}\left(-\phi+\gamma \right)}, \ z^*_1 &= \sqrt{r} \cos \left(\theta /2\right) e^{-\frac{i}{2}\left(-\phi+\gamma \right)}, \\ \nonumber \label{CP}
z_2 =& \sqrt{r} \sin \left(\theta /2 \right) e^{\frac{i}{2}\left(\phi+\gamma \right)}, \ z^*_2 &= \sqrt{r} \sin \left(\theta /2 \right) e^{-\frac{i}{2}\left(\phi+\gamma \right)} . 
\end{eqnarray}
If we now identify 
\begin{equation} \label{Hopf}
x^i = \bar{z} \sigma^i z,
\end{equation}
where $\sigma^i$ are the Pauli matrices, we recover $x^i$ expressed in terms of spherical coordinates $(r, \theta, \phi)$. This is a Hopf fibration, mapping points from $S^3$ in $\textbf{C}^2$ into points of $S^2$ in $\textbf{R}^3$. The angle $\gamma$, with the topology $S^1$, cancels out in this relation. 

Instead of considering any (square-integrable) function $\Phi(z,z^*)$, let us restrict ourselves only on those of the form $\Phi(x)$, with $x$ defined in \ref{Hopf}. This Hilbert subspace is the same as the usual Hilbert space of QM in $\textbf{R}^3$. Actually, action of the naturally defined operators is the same as well
\begin{eqnarray} \label{ope}
 \Delta \Phi(x) &=& \frac{1}{r} \{z^*_\alpha, \{z_\alpha,\Phi(x)\}\} = \partial_{x_i} \partial_{x_i} \Phi(x), \\ \nn
 \hat{x}_i \Phi(x) &=& x_i \Phi(x), \\ \nn
 \hat{V}_i \Phi(x)&\equiv &\frac{1}{2}\left[ \Delta, \hat{x}_i\right] \Phi(x) \\ \nn
 &=& - \frac{i}{2r}\sigma ^i_{\alpha \beta} (z^*_\alpha \partial _{z^*_\beta} + z_\beta \partial_{z_\alpha}) \Phi(x) \\ \nn
 &=& -i \partial_{x_i} \Phi(x), \\ \nn
 \hat{L}_i \Phi(x) &=& \frac{i}{2}\{ x_i, \Phi(x) \} = \varepsilon_{ijk} \hat{x}_j \hat{V}_k \Phi(x) .
\end{eqnarray}
When the Hilbert space is the same and so is the algebra of operators, there is not much left to differ (the norms coincide as well). Therefore, we conclude that this is a reformulation of the ordinary QM on $\textbf{C}^2$ instead of $\textbf{R}^3$.  

Usually, one can introduce magnetic monopoles in QM by using a vector potential $A_i$ singular on a (half-)line, see \cite{dirac}, or by using two regular potentials, see \cite{Yang}. We will now show that there is another way of doing so in the $\textbf{C}^2$ formulation of QM. 

The first step is to lower our restriction for the Hilbert space states. So far, we have only been considering those of the form $\Phi(x)$, where $x$ depended on $z,z^*$ as specified in \ref{Hopf}. As a result, the wave-functions always contained equal powers of $z$ and $z^*$ (since so does $x$). As we will show shortly, magnetic monopoles can be described using states of the form
\begin{equation} \label{kStates}
\Phi_\kappa(z,z^*) = \Phi(x) \cdot \xi_\kappa, \ \xi_\kappa = \left( \frac{z_1}{z^*_1} \right)^{\frac{\kappa - \delta}{4}} \left( \frac{z_2}{z^*_2} \right)^{\frac{\kappa + \delta}{4}},
\end{equation}
or expressed using \ref{euler}
\begin{equation}
 \xi_\kappa = e^{i \frac{\kappa}{2} \gamma} e^{i \frac{\delta}{2} \phi}.
\end{equation}
As the angles $\gamma, \phi$ are periodic (with a period of $4\pi$), for the wave-function to be uniquely defined we need $\kappa$ and $\delta$ to be integers. Note that $\Phi_\kappa \Phi^*_\kappa$ always contain the same powers of $z$ and $z^*$, therefore can be expressed as a function of $x$ and has a proper probabilistic interpretation in $\textbf{R}^3$. 

It can be shown that the operators defined the same way as in \ref{ope} act differently now, as functions $\Phi_\kappa$ do not depend only on $x$. One can easily prove (using that  $\{z_\alpha , . \} = -i \partial_{z^*_\alpha} $, $\{z^*_\alpha , . \} = i \partial_{z_\alpha} $ and the chain rule for derivatives) that
\begin{eqnarray}
\varepsilon_{ijk} \hat{x}_j \hat{V}_k \Phi_\kappa &=&\left( \hat{L}_i + \frac{\kappa}{2} \frac{\hat{x}_i}{r} \right)\Phi_\kappa, \\
\left[ \hat{V}_i, \hat{V}_j\right]\Phi_\kappa &=&   i\frac{\kappa}{2}  \varepsilon_{ijk} \frac{\hat{x}_k}{r^3}\Phi_\kappa ,
\end{eqnarray}
which are exactly the relations holding for a system containing a magnetic monopole of strength $\mu$ as derived, for example, in \cite{zwanziger}. To complete the identification we need to set $\kappa / 2 = \mu$. The Dirac quantization condition states that $\mu$ has to be a half-integer, but as in our case $\kappa$ is an integer, the identification is perfect. 

The smoking gun evidence for the monopole presence is the vector potential, which can be extracted as
\begin{equation} \label{comVA}
\hat{V}_j \Phi_\kappa = (- i \partial_{x_j} \Phi(x))\xi_\kappa + A_j \Phi_\kappa, \ \ \  A_j= - \frac{i}{2r\xi_\kappa} \sigma ^j _{\gamma \delta} z_\delta (\partial _{z_\gamma} \xi_\kappa).
\end{equation} 
With our choice of $\xi_\kappa$ the only nontrivial component (in spherical coordinates) is
\begin{equation}
A_\phi = \frac{\delta + \kappa \cos (\theta)}{2r \sin (\theta) },
\end{equation}
which leads to Coulomb-like magnetic field 
\begin{equation}
B_i = (\mbox{rot } \textbf{A})_i  = -\frac{\kappa}{2} \frac{x_i}{r^3} .
\end{equation}
It is obvious that the $\delta$ part is irrelevant and can be gauged away. However, it can also be kept and used to direct the Dirac (half-)string, as the choice of $\delta = \pm \kappa$ produces potential singular on the north and the south pole correspondingly. 

We conclude that generalized Hilbert space of states of QM formulated in $\textbf{C}^2$ instead of $\textbf{R}^3$ describes magnetic monopoles of any strength allowed by the Dirac quantization condition.
\section{Noncommutative quantum mechanics}
As we have seen, QM does not need $\textbf{R}^3$ to be defined. There is a great interest in so-called noncommutative (NC) spaces, that means spaces whose points cannot be distinguished below some fundamental scale. Similarly to ordinary QM, where the inability to measure exactly the position and the momentum of a particle is encoded in $[\hat{x}_i, \hat{p}_j] \neq 0$, such spaces, sometimes also referred to as quantum, are described by an NC relation
\begin{equation}
[x_i, x_j] \neq 0 .
\end{equation}
For spaces equipped with Poisson structure, we can use the means of canonical quantization. In the case of the space $\textbf{C}^2$, which was discussed in the previous section, it is done as follows:
\begin{equation}
\begin{array}{lcl}
\{ z_\alpha ,z^*_\beta \} = - i\,\delta_{\alpha\beta} & \rightarrow & [z_\alpha , z_\beta^* ] =  \lambda \delta_{\alpha \beta},\\ 
 \{ z_\alpha ,z_\beta \} = 0 & \rightarrow &  [z_\alpha , z_\beta ] = 0, \\
 \{ z^*_\alpha ,z^*_\beta \} = 0& \rightarrow &  [z_\alpha^* , z_\beta^* ] = 0 ,
\end{array} 
\end{equation}
where $\lambda$ is a constant with the dimension of length which describes the scale below which one cannot distinguish two close points of space. These quantized coordinates can be realized using two sets of creation and annihilation (c/a) bosonic operators $a_\alpha, a_\alpha^+, \alpha = 1,2$ satisfying the usual commutation relations $[a_\alpha, a^+_\beta] = \delta_{\alpha \beta}$ and $0$ otherwise. These act on an auxiliary Fock space spanned by normalized states
\begin{equation}
|n_1,n_2\rangle= \frac{(a^+_1)^{n_1}\,(a^+_2)^{n_2}}{
\sqrt{n_1!\,n_2!}}\ |0\rangle.
\end{equation}
We can use the relation \ref{Hopf} to carry this to three dimensions, creating three-dimensional, rotationally invariant noncommutative space $\textbf{R}^3_\lambda$ defined by 
\begin{equation}
[x_i, x_j] = 2 i \lambda \varepsilon_{ijk} x_k. 
\end{equation}
This is usually the starting point which can be also obtained by considering a sequence of concentric fuzzy spheres filling the entire three dimensional space (instead of considering just one of a fixed radius). 
The Cartesian NC coordinates and the radial coordinate $r$ are expressed using the c/a operators in a similar fashion to \ref{Hopf} as
\begin{equation}
x_i = \lambda a^+ \sigma^i a, r = \lambda ( a^+_\alpha a_\alpha+1),
\end{equation}
note that $x^2 = r^2 - \lambda^2$. 

Following the same steps as in the previous section, we can construct QM in this NC space. The wave-functions are now functions $\Psi(a,a^+)$ of the c/a operators and their scalar product is defined as
\begin{equation} \label{norm}
(\Phi,\,\Psi)\ =\ 4\pi \lambda ^2 Tr [ \Phi \,\hat{r}\, \Psi] , \ \ \ \hat{r} \Psi = \frac{1}{2}\left( r \Psi + \Psi r \right)\, ,
\end{equation}
where the weight in the norm was chosen to obtain the correct commutative limit. This defined the Hilbert space, operators on it can be also defined using the c/a operators, for example, the NC counter-parts to \ref{ope} are defined as
\begin{eqnarray} \label{ope}
 \hat{H}_0 \Psi &=&-\frac{1}{2 \lambda r} [a^+_\alpha, [a_\alpha , \Psi ]], \\ \nn
 \hat{X}_i \Psi &=& \frac{1}{2}\left(x_i \Psi +\Psi x_i\right), \\ \nn
 \hat{V}_i \Psi&\equiv &i\left[ \hat{H}_0 , \hat{X}_i\right] \Psi  \\ \nn
 &=& - \frac{i}{2r}\sigma ^i_{\alpha \beta} (a^+_\alpha [a_\beta, \Psi] - a_\beta [a^+_\alpha, \Psi] ) \\ \nn
 \hat{L}_i \Psi &=& \frac{i}{2\lambda}[ x_i, \Psi].
\end{eqnarray}
This way we can realize QM in a similar way as in the previous chapter. The only difference is that we are not able to construct states localized in a region of radius smaller than $\lambda$ and that the energy has an upper cut-off $E_{\mbox{max}} \sim \lambda^{-2}$. 
Preserving the rotational symmetry has many advantages, for example, the Hydrogen atom problem remains exactly solvable, see \cite{LRL}. Study of the velocity operators \cite{vel} revealed how the UV cut-off manifest itself in terms of higher symmetry ($SO(4)$ instead of the expected $SO(3)$). 
\section{Generalized states}
We can now continue in a similar fashion as in the commutative case, generalize the considered Hilbert space to introduce monopoles. The wave-functions now contain an unequal number of creation and annihilation operators
\begin{equation} \label{states}
\Psi_{\kappa'}(e^{-i\tau} a^+, e^{i\tau} a) = e^{-i\tau\kappa'} \Psi_{\kappa'} (a^+,a), \ \tau \in  \textbf{R}, \  \mbox{fixed}\  \kappa' \in \textbf{Z}  ,
\end{equation}
where $\kappa'$ counts their difference.

As was shown in \cite{mm}, the algebra of operators reproduces that of magnetic monopoles, but now with an occasional $\lambda$ dependent correction, for example
\begin{equation}
\left[\hat{V}_i,\hat{V}_j \right]  = i\frac{-\kappa'}{2} \varepsilon_{ijk} \frac{\hat{X}_k}{\hat{r}(\hat{r}^2-\lambda^2)} .
\end{equation}
In \cite{mm} it has been shown that the Coulomb problem of a system containing magnetic monopoles constructed this way remains exactly solvable. As it turned out in \cite{mm2}, most of the important components of QM (and possibly even more) are consistent with the monopole structure. If we denote $\hat{a} \Psi = a \Psi$ and $\hat{b}\Psi = \Psi a$ (and similarly for creation operators), then
\begin{eqnarray} \label{S}
\hat{S}_{ij} &=& \frac{1}{2}\,\varepsilon_{ijk}\,(\hat{a}^+\,\sigma_k\,\hat{a}\,-\, \hat{b}^+\,\sigma_k\,\hat{b}) ,\ \ \  \hat{S}_{k4}\ =\ \frac{1}{2}\,(\hat{a}^+\,\sigma_k\,\hat{a}\,+\, \hat{b}^+\,\sigma_k\,\hat{b})\,, \nn \\ 
\hat{S}_{05} &=& \frac{1}{2}\,(\hat{a}^+\,\hat{a}\,+\, \hat{b}^+\,\hat{b}) ,\hskip2cm \hat{C}\ =\ \hat{a}^+\,\hat{a}\,-\, \hat{b}^+\,\hat{b}\,, \\ \nn
\hat{S}_{0k} &=& \frac{i}{2}\,(\hat{a}^+\,\sigma_k\,\hat{b}\,-\, \hat{b}^+\,\sigma_k\,\hat{a})\,, \ \ \ \hat{S}_{45}\ =\ \frac{i}{2}\,(\hat{a}^+\,\hat{b}\,-\, \hat{b}^+\,\hat{a})\,,\\ 
\hat{S}_{k5} &=& \frac{i}{2}\,(\hat{a}^+\,\sigma_k\,\hat{b}\,+\, \hat{b}^+\,\sigma_k\,\hat{a})\,, \ \ \ \hat{S}_{04}\ =\ \frac{1}{2}\,(\hat{a}^+\,\hat{b}\,+\, \hat{b}^+\,\hat{a})\, ,
\end{eqnarray}
is an operator representation of $su(2,2)$ accompanying the velocity operator, the angular momentum operators, the free Hamiltonian, the dilation operator and the Laplace-Runge-Lenz operator (in some cases the factor of $\hat{r}^{-1}$ is needed to assure Hermiticity under \ref{norm}, see the discussion in \cite{mm2}). Action of the central element $\hat{C}+2$ is on $\Psi_{\kappa'}$ states equal to $\kappa'$. In other words, the way of describing magnetic monopoles in NC QM by considering a generalized class of physical states is consistent with the most important physical operators already present.
\section{Conclusions}
We have presented two different topics. The first was an alternative description of magnetic monopoles in quantum mechanics, the other was a construction of quantum mechanics in an NC space. Both of them utilized the close relation between the spaces $\textbf{C}^2$ and $\textbf{R}^3$. It is the extra degree of freedom that allows the description of monopoles and it is the natural Poisson structure that allows the quantization of the space(s). Therefore, it shall be of no surprise that both of these are compatible and NC QM with magnetic monopoles can be constructed as well. 

Even though we find the succession presented in this report more logical, the historical order is that we analyzed NC QM first, see \cite{vel, LRL, plane, mbh}, then found out the generalization introducing the monopoles, see \cite{mm,mm2}, and considered the commutative theory only afterward, see \cite{mmc}. The outlook, for now, is to utilize the found symmetries and consider the relativistic generalization of the theory.

\section*{Acknowledgments}
This research was partially supported by COST Action MP1405 (S.K. and P.P), project VEGA 1/0985/16 (P.P.) and the Irish Research Council funding (S.K.).  The speaker is grateful for the hospitality provided to him at the Corfu Summer Institute 2017.

\end{document}